\documentclass[prl,twocolumn,floatfix,amsmath,amssymb]{revtex4}
\pdfoutput=1
\usepackage{graphicx}
\usepackage{dcolumn}
\usepackage{bm}
\usepackage{amssymb}
\usepackage{amsmath}
\usepackage{color}

\usepackage{times}

\begin{document}

\title{Hidden order revealed in quantum oscillations in cuprate superconductors}

\author{Xun Jia}
\author{Ivailo Dimov}
\author{Pallab Goswami}
\author{Sudip Chakravarty}
\affiliation{Department of Physics and Astronomy, University of
California Los Angeles, Los Angeles, California 90095}

\date{\today}

\begin{abstract}
We follow the line of reasoning that hidden broken symmetries are the root of quantum oscillations  observed 
in underdoped superconductors and examine the role of bilayer splitting and incommensuration. This is a
view that eschews the notion of a featureless Mott liquid as the source of complexity. Instead, our view 
is grounded in a conventional Fermi surface and quasiparticles. We show that bilayer splitting and/or 
incommensurate $d$-density wave order can lead to many interesting results, in particular a splitting of the main frequency of the quantum oscillations.

\end{abstract}


\maketitle 
 \paragraph{Introduction:} Over the past twenty years of the discovery of the high temperature superconductors, the imperfections of the materials have hidden many important facts and given rise to a set of dogmas that are all pervasive. Inspired by the recent quantum oscillation experiments in underdoped cuprates, we depart from the accepted wisdom and place prominence to the notion of broken symmetries~\cite{Chakravarty:2008}. Broken symmetries associated with phases of matter dictate the elementary excitations that determine the macroscopic properties, which offer us clear clues about the gross features of a physical problem and transcends whether or not the system is strongly correlated. It is this view that inspired us to propose the picture of the $d$-density wave (DDW) that can potentially unify a wide range of disparate features of the cuprates~\cite{Chakravarty:2001}. 
 
The quantum oscillation experiments~\cite{Doiron-Leyraud:2007,Bangura:2008,LeBoeuf:2007,Jaudet:2008,Yelland:2008} indicate that in highly underdoped cuprates: (1) the  de Haas-van Alphen (dHvA) effect, the Shubnikov-de Haas (SdH) effect, and the oscillations of the Hall coefficient, all have the same underlying cause, the conventional quantization of the Landau levels, meaning oscillations that are periodic in the variable $1/B$, where $B$ is the magnetic induction; (2) the Hall coefficient in high magnetic fields and low temperatures is negative; (3) given the conventional nature of the oscillation spectra, it is difficult to avoid the Luttinger theorem~\cite{Luttinger:1960,Dzyaloshinskii:2003} regarding the volume of the Fermi surface and the number of charge carriers; (4) there are strong indications that Fermi surface reconstruction due to a broken symmetry plays an important role; (5) there are also tantalizing hints that order may even incommensurate~\cite{Sebastian:2008};  if there is incommensurate DDW order, this will strongly enhance the tendency to inocmmensurate charge density wave order (CDW) with twice the ordering vector~\cite{Zachar:1998}. Thus, indirectly through the intermediate CDW, there may be a link between the incommensurabilities of the spin density wave (SDW) and DDW; (6) dHvA in the mixed state can reflect the properties of the normal state. In a recent paper, one of us has addressed some of these aspects from the perspective of DDW~\cite{Chakravarty:2007}.

Here we shall concentrate on the following: (1) the effect of bilayer splitting in $\mathrm{YBa_{2}Cu_{3}O_{6+\delta}}$ (YBCO), an unavoidable structural constraint that one must address; (2) possible incommensuration of the order parameter~\cite{Chakravarty:2001,Dimov:2005}; (3) the robustness of dHvA in the mixed state reflecting the putative normal state; (4) some important unanswered questions. Our main analysis is based on a Hartree-Fock theory of the order parameter and the Gorkov equations for the mixed state. It is the beauty of an order parameter theory that those features that are protected by symmetries can be explored in the weak interaction limit, although the system may be  strongly interacting.

\paragraph{Bilayer splitting}
Bilayer coupling, $t_{\perp}({\bf k})$, in YBCO is well-known. It has been parametrized in terms of a momentum conserving tunneling matrix element. For tetragonal structure it is~\cite{Chakravarty:93,Andersen:1995}
\begin{equation}
t_{\perp}({\bf k})=\frac{t_{\perp}}{4}\left[\cos (k_xa)-\cos (k_ya)\right]^2,
\label{eq:tmatrix1}
\end{equation}
where $a$ is the lattice spacing.

The outline of the calculation is as follows. The total Hamiltonian 
$
H=H_{1}+H_{2}+H_{12},
$
where $H_{1}$ and $H_{2}$ are the full Hamiltonians of the layers $(1)$ and $(2)$ and their spectra are degenerate, as they are identical. The tunneling Hamiltonian $H_{12}$ is given in the momentum conserving case to be
\begin{equation}
H_{12}=\sum_{{\bf k},\sigma}t_{\perp}({\bf k})(c_{{\bf
k},\sigma}^{\dagger (1)}c_{{\bf
k},\sigma}^{(2)}+1\leftrightarrow 2)
\end{equation}

\paragraph{Commensurate order:} We now argue that in the Hartree-Fock approximation, $H_{0}=H_{1}+H_{2}$ can be written
as the effective DDW Hamiltonians
\begin{eqnarray}
H_{0}&=&\sum_{{\bf k} \in RBZ,\sigma}(\epsilon_{\bf k}c_{{\bf
k},\sigma}^{\dagger (1)}c_{{\bf
k},\sigma}^{(1)}+\epsilon_{\bf k+Q}c_{{\bf
k+Q},\sigma}^{\dagger (1)}c_{{\bf
k+Q},\sigma}^{(1)}+1\leftrightarrow 2)\nonumber \\ 
&+&\sum_{{\bf k} \in RBZ,\sigma}(i \, W_{\mathbf{k}} c_{\mathbf{k} \sigma}^{\dagger (1)} c_{\mathbf{k}+\mathbf{Q},\sigma}^{(1)}+1\leftrightarrow 2+\text{h. c.}) 
\label{eq:heff}
\end{eqnarray}
where ${\bf Q}=(\pi/a,\pi/a)$, and  $\epsilon_{\bf k}$ is the single particle spectra. The superscripts on the electron creation and annihilation operators stand for the layer index. The reduced Brillouin zone (RBZ) is bounded by $k_{y}\pm k_{x}=\pm \pi/a$. We shall parametrize $\epsilon_{\bf k}$ by~\cite{Andersen:1995}
\begin{eqnarray}
 \epsilon_{\mathbf {k}}&=& - 2 t (\cos k_{x}a+\cos k_{y}a)+4 t' \cos k_{x}a\cos k_{y}a\nonumber \\ 
 &-&2t'' (\cos 2 k_{x}a+\cos 2 k_{y}a)
\end{eqnarray}
and the DDW gap  by
$W_{\bf k}=\frac{W_{0}}{2}(\cos k_{x}a-\cos k_{y}a).
$
With the choice of the quadratic Hamiltonian $H_{0}$ in Eq.~(\ref{eq:heff}), it can be easily diagonalized along with $H_{12}$. This is a first order  degenerate perturbation theory. Because $t_{\perp}$ will turn out to be so small, we do not expect a large correction.  

At each wave vector $k$ in the RBZ, we need to diagonalize a $4\times 4$ matrix to extract the energy eigenvalues. This matrix 
 is
\begin{equation}
\mathbb{H} = \left(\begin{array}{cccc}\epsilon_{\bf k} & iW_{\bf k} & t_{\perp} (\bf k)& 0 \\- iW_{\bf k} & \epsilon_{\bf k+Q}  & 0 & t_{\perp} (\bf k+Q) \\t_{\perp} (\bf k)& 0 & \epsilon_{\bf k} &  iW_{\bf k} \\0 & t_{\perp} ({\bf k+Q}) & - iW_{\bf k}  & \epsilon_{\bf k+Q} \end{array}\right).
 \label{eq:h}
\end{equation}
The energy eigenvalues are, with $s=\pm 1$,
\begin{equation}
E_{s}^{\pm}({\bf k})=\frac{\epsilon_{\bf k}  + \epsilon_{\bf{ k+Q}}-2s t_{\perp}({\bf k})}{2}\pm \frac{\sqrt{(\epsilon_{\bf k}- \epsilon_{\bf k+Q})^2+4
   W_{\bf k}^2}}{2}
 \end{equation}
 
 From $t_{\perp}({\bf k})$ it is clear that the electron pockets will be much more affected by it than the hole pockets. The bilayer splitting of the main frequency in the dHvA meausurement, presumed to be from the electron pockets, if it is to occur, should be smaller or of the order of the half width at the half maximum of the peak in the Fourier spectra, otherwise it would have been already resolved~\cite{Jaudet:2008,Sebastian:2008}. This fact combined with the Luttinger sum rule strongly constrains its magnitude.

\begin{figure}
    \centering
  \includegraphics[width=7.5cm]{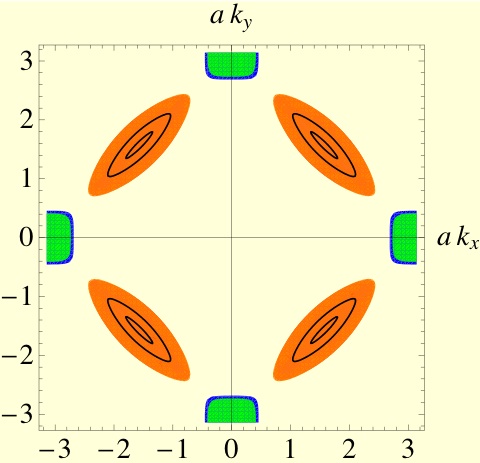}\\
  \caption{(Color online)Fermi surface split by bilayer coupling for commensurate DDW. The hole pockets centered at $\frac{1}{2}(\pm\pi/a,\pm\pi/a)$ are essentially unsplit by the bilayer coupling, In contrast, the electron pockets centered at  $(\pi/a,0)$ and symmetry related points are split as described in the text.}
  \label{fig:C}
\end{figure}

 The parameters we choose  for YBCO at 10\% doping are~\cite{Chakravarty:2007}: $t = 0.3 \; \text{eV}$, $t' = 0.3 t$, $t''=t'/9.0$, $t_{\perp}=8\; \text{meV}$, and $W_{0}=0.0825\; \text{eV}$. With these choices of the parameters and the chemical potential $\mu$ set to $-0.2627\; \text{eV}$, we get the total hole doping of $n_{h}\approx 10\%$, constrained by the Luttinger sum rule~\cite{Luttinger:1960}, a  {\em sine qua non} of the Fermi liquid picture. The corresponding dHvA frequencies according to the Onsager formula  are $F_{1}\approx 944\; \text{T}$, $F_{2}\approx 967\; \text{T}$, $F_{3}\approx 570\; \text{T}$, and $F_{4}\approx 450\; \text{T}$. The frequencies $F_{1}$ and $F_{2}$ correspond to the hole pockets and are essentially the same within our accuracy, while $F_{3}$ and $F_{4}$ correspond to the electron pockets split by the bilayer coupling. The Fermi surfaces are shown in Fig.~\ref{fig:C}.

\paragraph{Incommensurate order:}
In view of a recent experiment~\cite{Sebastian:2008}, it is interesting to pursue what incommensurate DDW order should predict for dHvA.
Mean field theory with a class Hamiltonians show that it is difficult to achieve incommensuration, unless the $d$-wave order parameter in the particle-hole channel is mixed with a $s$-wave component (may be natural in the presence of orthorhombicity) or a $id_{xy}$ component~\cite{Dimov:2008}. 
The reason is that  it is energetically favorable to add holes at the nodes, unlike SDW, and provides a natural rigid band shift forming hole pockets and the consequent electron pockets from the band folding. Nonetheless, it is possible that the Fermi surface could eventually move away from the nesting wave vector as a function of doping~\cite{Chakravarty:2001,Kee:2002}. Whether or not this happens before the DDW gap collapses with increased doping  is a very difficult question to answer. Any incommensurate order is generally complex, especially if the incommensuration is irrational (a slight abuse of terminology). Nonetheless, a useful approximation is to keep  the largest gap and the hierarchy of gaps can be washed out due to thermal fluctuations, disorder, or magnetic breakdown~\cite{Falicov:1967}. A simple approximation~\cite{Dimov:2005} is given by the particle hole condensate
$
\langle c^{\dagger}_{\sigma'\mathbf k'} c_{\sigma\mathbf k}\rangle = i \frac{W_{\mathbf k}}{2} \left(\delta_{\mathbf {k', k+K}}+\delta_{\mathbf {k', k-K}}\right) \delta_{\sigma,\sigma'},
$
where the incommensuration vector ${\mathbf q}= {\mathbf K}-(\pi/a,\pi/a)$. This {\em Ansatz} conserves current to only quadratic order in $\mathbf q$, which may be sufficient for practical situations. 

\begin{figure}
    \centering
  \includegraphics[width=7.5cm]{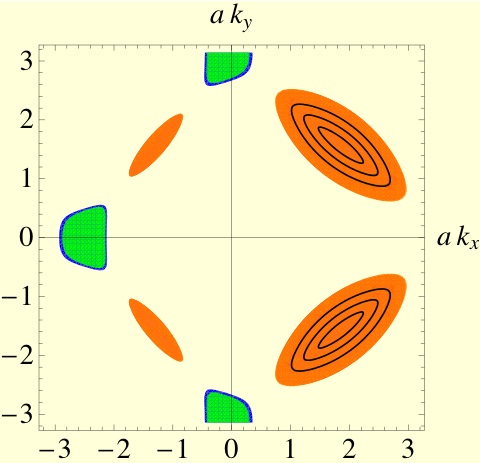}\\
  \caption{(Color online)Fermi surface split by bilayer coupling for incommensurate DDW. The hole pockets centered at $\frac{1}{2}(\pm\pi/a,\pm\pi/a)$ are essentially unaffected by the bilayer coupling, The electron pockets  are split as described in the text. The band structure parameters are the same as in Fig.~\ref{fig:C}; $W_{0}=0.115\; \text{eV}$,  $t_{\perp}=6 \; \text{meV}$, $\eta=0.09$, $\mu=-0.2575 \; \text{eV}$, and $n_{h}\approx 10\%$, as before. The chosen order parameter spontaneously breaks time reversal symmetry.  Thus, when reflection symmetry is also broken, an asymmetric band structure is not only allowed, but is expected on general grounds. }
  \label{fig:IC}
\end{figure}

There is some evidence from neutron scattering of incommensurate SDW fluctuations  ${\mathbf q}= \pi(\pm  2\eta,0)/a$ and ${\mathbf q}= \pi(0,\pm  2\eta)/a$, where $\eta \sim 0.1$ in underdoped YBCO~\cite{Dai:2001}. A similar  estimate should also apply to DDW because the determining competition between the kinetic and the interaction energies are similar within a mean field theory; the precise value of $\eta$ is not particularly important at this time. We further assume  a single wave vector $\mathbf q$ by spontaneous breaking of inversion symmetry, guided by the experimental observations. Note, however, that the product of inversion and time reversal, which is broken by DDW, is preserved, and we do not expect macroscopic currents. (One could have taken the order to have wave vectors $\bf q$ and $-\bf q$, and thereby not breaking inversion symmetry, resulting in a different Fermi surface.) The excitation spectrum can be trivially solved. The Hamiltonian is the same as in Eq.~\ref{eq:h}, except that the wave vector $\bf k$ now runs over the full Brillouin zone and $W_{\bf k}$ must be replaced by $W_{\bf k}\to \frac{1}{2}(W_{\bf k}-W_{\bf k-K})$ and ${\bf Q}$ by ${\bf Q}\to {\bf K}$. The resulting Hamiltonian is easily diagonalized and the Fermi surfaces are shown in Fig.~\ref{fig:IC}. The corresponding dHvA frequencies according to the Onsager formula  are $F_{1}\approx 1661\; \text{T}$, $F_{2}\approx  251\; \text{T}$, $F_{3}\approx  535\; \text{T}$, and $F_{4}\approx  442\; \text{T}$. The frequencies $F_{1}$ and $F_{2}$ corresponding to the hole pockets are essentially unchanged by the bilayer coupling, whereas $F_{3}$ and $F_{4}$ corresponding to the electron pockets are split by the bilayer coupling. 

\paragraph{Robustness of dHvA in the mixed state:}
Many superconductors exhibit dHvA effect deep in the mixed state (fields as low as half the upper critical field) with  frequencies unshifted from the putative normal Fermi liquid state,  except that the amplitude of the oscillations is diminished~\cite{Wasserman:1996}. A particularly interesting analysis was given by Stephen~\cite{Stephen:1992}. To show that the same is true if the normal state is the DDW with a reconstructed Fermi surface,  it is sufficient to examine the hole pockets close to the nodal points where the Hamiltonian is given by an effective Dirac Hamiltonian, as  the electron pockets are described by non-relativistic fermions and Stephen's results can be immediately taken over. The nodal Hamiltonian  in the Landau gauge is~\cite{Yang:2002}
$
H_{D}=\sqrt{v_Fv_D}\left[p_{\tilde{x}}\hat{\sigma}_3+(p_{\tilde{y}}-eB\tilde{x}/c)\hat{\sigma}_2\right],
$
where $\hat{\sigma}$'s are the standard Pauli matrices and the spatial anisotropy was   removed with the redefinition of the coordinates $\tilde{x}=\sqrt{v_D/v_F}x$ and $\tilde{y}=\sqrt{v_F/v_D}y$, where $v_{F}$ is the velocity perpendicular to a constant energy contour and $v_{D}$ is the velocity tangential to it. The momenta should similarly be rescaled, and we denote them by $\tilde{k}$. The unperturbed Green's function resulting from this Hamiltonian is used to solve the Gorkov equations~\cite{Abrikosov:1975} to find a self energy matrix, $\Sigma_{n_{1}, k_{1},\alpha_{1}; n_{2}, k_{2},\alpha_{1}}(i\omega)$, which being diagonal in the Landau index and the wavevector (also independent of it), is, assuming that the Landau level index $n\gg 1$ (the full expression can be found in Ref.~\cite{Goswami:2008}), 
\begin{eqnarray}
\Sigma_{n,\alpha_1,\alpha_2}(i\omega)&\approx&\frac{\Delta^2 \mu\sqrt{\pi}}{4\sqrt{n}(\hbar \tilde{\omega})^2}\left[\frac{1}{\sqrt{\pi n}}\frac{\mu^2-\epsilon_{n}^{2}}{(\hbar \tilde{\omega})^2}-i\ \textrm{sgn}(\omega \mu)\right]\nonumber \\
&\times& (1-\alpha_{1})(1-\alpha_{2})
\end{eqnarray}
Here $\epsilon_{n,\tilde{k},\alpha}=\alpha \sqrt{n} \hbar \tilde{\omega}=\alpha\epsilon_{n}$, $\tilde{\omega}=\sqrt{2}v_{F}v_{D}/\ell$, and $\ell=\sqrt{\hbar c/e B}$; $\alpha=-1$ for the lower branch of the Dirac spectra and  $\alpha=+1$ for the upper branch . For simplicity we have used an $s$-wave superconducting gap $\Delta$, but it is straightforward to generalize to a $d$-wave gap. The main result is unchanged, namely  that the real part is very small compared to the imaginary part. (An interesting aspect of this formula is that only for the component $\alpha_{1}=\alpha_{2}=-1$ the result is non-vanishing.) Thus, we can define an ``impurity''  scattering rate at the Fermi energy due to vortices in the mixed state as 
\begin{equation}
\frac{\hbar}{\tau_{v}}=\sqrt{\frac{\pi}{8}}\frac{\Delta^2}{\sqrt{\mu \hbar \omega_{c}^{*}}},
\end{equation}
where the effective mass for the DDW quasiparticles is defined by $m^{*}=\mu^2/(v_Fv_D)$ and $\omega_{c}^{*}=eB/m^{*}c$. In agreement with Stephen, the cyclotron motion is so much faster than the vortices that the electrons see the vortices as static impurities. It is now obvious that there would be an additional Dingle factor arising from the vortices in the mixed state, but the dHvA frequencies will be unshifted to an excellent approximation from the putative normal DDW state; the superconducting gap sets only the magnitude of $\tau_{v}$. A more complete discussion will be given in Ref.~\cite{Goswami:2008}. It is not difficult to see, however, that for a $d$-wave superconducting gap $\Delta^{2}$ will be replaced by the Fermi surface average of the square of the $d$-wave superconducting gap.

\paragraph{Coda:} There are important questions that remain unanswered. The most glaring  is the Fermi surface determined in angle resolved photoemission spectroscopy (ARPES), which is mostly consistent with Fermi arcs~\cite{Norman:1998} in the relevant doping regime in the normal state. It is noteworthy, however, that in electron doped cuprates, the evidence of hole and electron pockets is a solid but an often forgotten fact~\cite{Armitage:2001}. More recent  ARPES experiments have indicated remarkable data for hole pockets~\cite{Chang:2008}. In contrast, experiments in YBCO films where the doping is adjusted by applying potassium overlayer to reduce it to 10\% see once again a Fermi arc~\cite{Hossain:2008}. Although by itself this is not in contradiction with the notion of a hole pocket because it has long been argued that the back side of the hole pocket should have very small intensity in ARPES simply from the DDW coherence factors~\cite{Chakravarty:2003}. The real issue is of course why the electron pockets are not seen, at least the two sides of it. Perhaps adding potassium overlayer may cause significant charge scattering and the electron pockets may be highly fragile with respect to disorder.

One might argue that the ortho-II potential of YBCO6.5 might be important~\cite{Bascones:2005}, a material for which data for Hall, SdH, and dHvA data exist. We believe that this may not be so because highly polarizable BaO-layers next to the chains should screen the potentials quite effectively, to the extent that even disordered chains appear to have little effect in the planar physics in many properties. It is also important to recognize that SdH  (and dHvA) measurements are also available in  $\mathrm{YBa_{2}Cu_{4}O_{8}}$, a double chain compound, and even negative Hall coefficient is clearly observed in agreement with the ortho-II materials. It is difficult to believe that this universality is achievable if the chain potentials were playing a significant role. Note, however, that as yet no oscillations are observed in ortho-VIII,  although negative Hall coefficient  is similarly observed~\cite{LeBoeuf:2007}.

 We are fully cognizant  that neutron scattering experiments have resulted in a controversy, as commented in Ref~\cite{Chakravarty:2007}, with regard to  DDW. The fact remains that if the pseudogap is supposed to reflect an order with a strongly momentum dependent gap  in the particle-hole channel, this is impossible without breaking translational symmetry.
  
Bilayer splitting in overdoped materials is  well established~\cite{Feng:2001}, although the actual splitting, $88\; \text{meV}$, is dramatically smaller than the local density band structure calculations, which is $300\; \text{meV}$. The reason for this is likely to be the $Z$-factor associated with the Green function~\cite{Abrikosov:1975}. We have argued that the splitting at doping of 10\% is even smaller and only  about 12-16 meV. If this is the case, a more precise  understanding of this renormalization is necessary, but it could not be  a breakdown of the Fermi liquid theory, as these oscillation measurements are indicative of quasiparticles of a Fermi liquid.

So far  the high frequency peak at 1654 T  observed in DC measurements~\cite{Sebastian:2008} has not been observed  in pulsed field  measurements~\cite{Jaudet:2008}. The commensurate order, as we have seen, predicts three main frequencies, two of which are from bilayer split electron pockets, but the hole pocket remains essentially unsplit at about 970 T; however,  this hole pocket frequency is not yet seen (although there is a very weak signature in some data), may be because it lies too close to the second harmonic of the electron pocket frequency. On the other hand, if the order is incommensurate, the observed frequency at about 530 T (electron pocket) should be similarly split by the bilayer coupling. The hole pocket frequency at about 1650 T will essentially remain unsplit because of the nature of the bilayer coupling.  To conserve the Luttinger sum rule, however, there should be a lower hole frequency at about 250 T, whose splitting will be similarly very small.

Although there are other analyses of the quantum oscillation phenomenon~\cite{Lee:2007,Millis:2007,Kaul:2007,Chen:2008}, the prediction of the splitting of the main frequency due to bilayer coupling and a smaller hole pocket frequency at 250 T for the incommensurate case are entirely new, and, if verified, would constitute a striking argument for the Fermi liquid picture---for a doping, as low as 10\%, this would be surprising to say the least.

This work is supported by NSF under Grant No. DMR-0705092. We thank S. Sebastian, C. Proust, S. Kivelson, and L. Taillefer for comments and correspondence. We also thank R. B. Laughlin and C. Nayak for encouragement.

\paragraph{Note added:} After our work was completed, we received a complementary paper~\cite{Podolsky:2008}. It emphasizes ortho-II ordering  that results in a very different Fermi surface.  We have argued above  that ortho-II potential is likely to be unimportant. The treatment of bilayer coupling in Ref.~\cite{Podolsky:2008} does not yield the splitting that we have emphasized. Moreover, neither the incommensuration nor the analysis of the mixed state is discussed there.

\end{document}